\documentclass{article}
\usepackage{graphicx}
\usepackage[legalpaper, margin=1.25in]{geometry}
\usepackage{authblk}

\usepackage{lineno}

\usepackage{amsfonts}
\usepackage{amsmath}
\usepackage{amssymb}
\usepackage{amsthm}
\usepackage{algorithmic}
\usepackage{algorithm}
\usepackage{bm}
\usepackage{quantikz}
\usepackage{mathastext}
\usepackage{mathtools}
\usepackage{physics}
\usepackage{bbm}
\usepackage[sortcites,sorting=none,backend=biber]{biblatex}
\usepackage{multicol}
\usepackage{placeins}
\usepackage{caption}
\usepackage{subcaption}
\usepackage{dsfont} 
\usepackage{hyperref}
\usepackage[capitalize,noabbrev]{cleveref}

\crefname{paragraph}{Subsection}{Subsections}

\makeatletter
\renewcommand\paragraph{\@startsection{paragraph}{4}{\z@}%
{-3.25ex\@plus -1ex \@minus -.2ex}%
{1.5ex \@plus .2ex}%
{\normalfont\normalsize\bfseries}}

\begin{document}

\tikzset{every picture/.style={line width=0.75pt}} 

\title{Threat Vectors and the State of the Art in Defense Methods for Security in Neurotechnology}
\small
\author[1,2]{Bryce Allen Bagley}
\author[1,3]{Nathaniel Rose}
\author[4]{Quintus Kilbourn}
\author[1,2]{Matthew Canham}

\affil[1]{Cerberus Neurosecurity Research Institute}
\affil[2]{Cognitive Security Institute}
\affil[3]{Dura Labs}
\affil[4]{Flashbots}
\maketitle              

\begin{abstract}
    Brain-computer interfaces (BCIs) are a class of diverse hardware modalities, associated software, and connected devices which are widely used in a variety of fields, including neurosurgery, biomedical data analysis, and neuroimaging. Recent years have seen rapid advancements in BCI technology, and neurotechnology more broadly, with the first devices now passing clinical trials, early examples of consumer hardware entering the market, and many variants of consumer and medical hardware with increasingly extensive capabilities being developed rapidly. However, research and development in security for BCIs--known as neurosecurity--lags significantly behind the capabilities of BCIs themselves. In an effort to address as many vulnerabilities as feasible immediately, in this paper we review the current state of the art in neurosecurity, thoroughly survey the breadth and complexity of both firmly established and highly probable security threats to BCI systems, and provide recommendations of existing methods from cybersecurity, hardware security, and machine learning which can immediately be applied to address some of these gaps in neurosecurity.
\end{abstract}

\section{Introduction}

Recent years have seen rapid advancement in brain-computer interfaces (BCIs) and other neurotechnologies, particularly within the realm of systems capable of translating thought into speech, text, or device commands.\cite{Pandarinath_2017_high_performance_bci_speech_paralyzed_patients,Roelfsema_2018_mind_reading_and_writing_future_of_neurotech,Willett_2021_high_performance_BCI_brain_to_text_communication,Willett_2023_high_performance_speech_neuroprosthesis,Drew_2023_rise_of_brain_reading_technology,scotti_2024_transfer_learning_MRI_brain_readout} However, these expanding capabilities have brought with them a growing awareness of associated privacy and security risks. 

In the decade and a half since the first paper on neurosecurity,\cite{Denning_2009_neurosecurity_first_paper_neurosurgery_journal} a number of researchers have attacked problems in this space. EEG systems have been a primary focus,\cite{Landau_EEG_based_bci_security,meng2023eegbackdoor,Belkacem_cybersecurity_P300_BCI}, but a number of papers have described various attack types,\cite{Lange_2018_BCI_attack_pin_number_recovery,bernal_2020_brain_implant_neuronal_cyberattacks_experiments_1,armengolurpi2023brainhack,professorx2024} explored information-theoretic questions related to neurosecurity,\cite{bonaci_2014_brain_security_exocortex,bagley_petritsch_2024_holographic_cognition_original_cognitive_neurosecurity}, or reviewed a range of known and anticipated issues in neurosecurity.\cite{takabi_2016_bci_privacy_threats_review,PYCROFT_2016_brainjacking_bci_security,canham_sawyer_2019_neurosecurity_MASINT,bernal_2022_bci_security_review,Xia_2023_privacy_preserving_BCI_review} However, compared with the scope of the threats and the pace at which neurotechnology is advancing, concerningly little research has yet been done on neurosecurity. Additionally, the spectrum of attack surfaces for neurotechnology--especially BCI systems--has not been fully explored in any of these reviews, nor has there been a comprehensive review of the wealth of techniques available from other fields which can immediately be applied to address certain threats. To our knowledge the review by Bernal et al comes closest to the former task,\cite{bernal_2022_bci_security_review} and due to the slow pace of progress in the field it is still very relevant. However, in this paper we assess a number of attack surfaces which are absent from that review, and additionally our focus is on immediately actionable solutions rather than solely a survey of the state of the art in neurosecurity broadly. 

Our goal in this paper is to provide as useful a review as possible for the engineers, scientists, and developers creating BCIs, as well as provide relevant technical insight for those concerned with developing effective neural data policy grounded in the reality of contemporary technology. By drawing upon the expertise of a highly interdisciplinary team spanning neuroengineering, cognitive science, computer engineering, hardware security, systems science, biological physics, and medicine, we have endeavored to compile as comprehensive as possible a taxonomy of attack surfaces relevant to neurotechnology, ranging from threats to supply chains and ASIC-level hardware all the way to cloud servers and machine learning models, and from neurophysiology to cognitive phenomena. We place an additional emphasis on examples of how targeting multiple attack surfaces simultaneously can create new attacks which may be more potent, harder to mitigate and detect, or both. Further, we provide an extensive sampling of the literature on methods from cybersecurity, hardware security, robust and private machine learning, software methods, and more which are immediately applicable for practitioners.

The paper is organized into two primary sections. We begin with our taxonomy of attack surfaces for neurotechnology, and in many cases include references to methods for mitigating such attacks. In the second portion of the paper, we discuss additional defense methods from relevant fields which we felt did not cleanly fit into discussions of any specific attack surface, as well as methods which were relevant to multiple attack surfaces in order to avoid redundancy in the first section.

The nature of security research is to evolve over time. As systems, formats, standards, and other technologies change, some methods become outdated and new challenges and approaches arise. This review should be treated as reflective of a single snapshot in time of neurosecurity methods, early in what appears to be an emerging revolution in neurotechnology. Some principles discussed herein are almost certain to persist indefinitely. For example, the necessity of post-quantum cryptography, the threat of air-gap attacks, and the central importance of considering neurophysiology and cognition as attack surfaces. Others may change, and so those assessing security threats and defense methods in the future should ensure they are basing design choices on the most up-to-date information.

\begin{figure}
    \centering
    \includegraphics[width=0.75\linewidth]{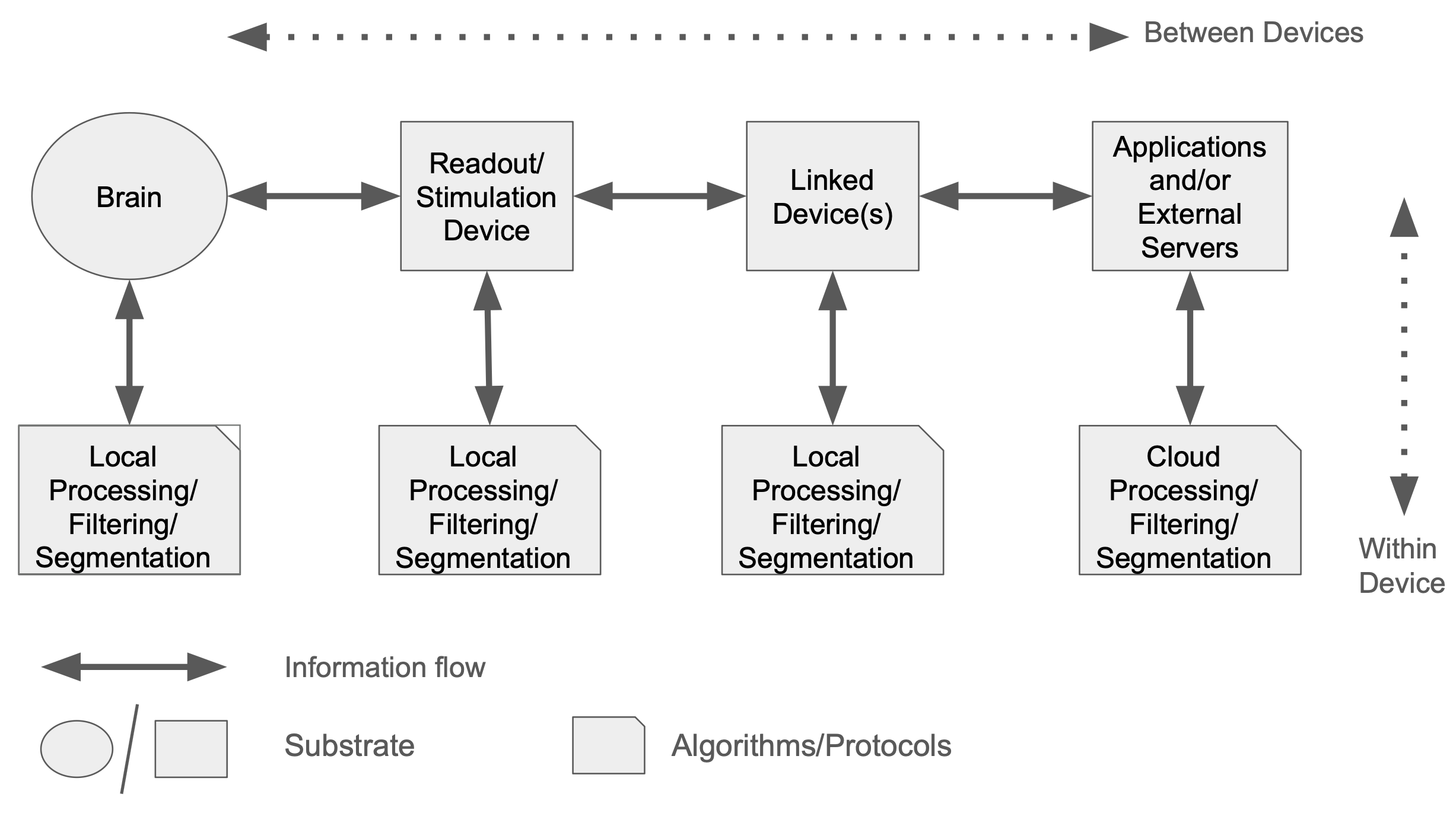}
    \caption{A simplified visualization categorizing the attack surfaces in the BCI cycle. While different attacks are relevant at different points, each arrow and shape represent a respective point of vulnerability where one or more attacks are relevant.}
    \label{fig:BCI_cycle_unrolled}
\end{figure}

\section{Attack Surfaces and Attack Types}\label{section:attack_surfaces_and_types}

The BCI cycle provides a helpful conceptual structure for assessing the attack surfaces present in any system which incorporates a BCI.\cite{bernal_2022_bci_security_review} The typical diagrams depict it in a circular form, but the resultant visual compression can sometimes obfuscate certain attack surfaces. As such, here we use a modified diagram in which the cycle is "unrolled", so to speak, into a linear form. This can be seen in \cref{fig:BCI_cycle_unrolled}.

In \cref{fig:BCI_cycle_unrolled}, each arrow and shape reflects an attack surface (or a collection of them) which an adversary might attempt to target. These include more traditional attack surfaces such as wireless communications, cloud services, end use applications, and more, as well as more exotic and BCI-specific surfaces such as the physical substrate of the brain, machine learning models operating with neural data, and the functional processes brains engage with, ranging from cognition to motor, endocrine, or autonomic signaling. More traditional attack surfaces can easily benefit from the ability to directly or relatively directly adopt existing security methods, such as encryption, secure communications protocols, differential privacy, and more. However, the closer to the brain and neural data one is operating--in terms of both literal and metaphorical distance--the more likely there will be a need to modify these protocols or develop entirely new ones. Examples of these more difficult problems include human cognition and differences in the statistical and computational properties of neural function vs more traditional computing.\cite{bagley_petritsch_2024_holographic_cognition_original_cognitive_neurosecurity,bagley_2025_approximatingmathematicalstructurepsychodynamics} However, there is already at least one formal protocol for targeted obfuscation of neural data outputs from a BCI to enhance privacy,\cite{bonaci_2014_brain_security_exocortex,Ienca_2018_brain_leaks_cybersecurity_first_mention_of_differential_privacy} and Bagley and colleagues are working on protocols for cognition.\cite{bagley_2025_aria_talk}

\subsection{Common Threat Modalities}

There are a number of threats to neurosecurity which apply across much of, if not the entirety of, the BCI cycle. Examples include threats such as "Harvest Now, Decrypt Later" (HNDL) attacks, air gap attacks, man-in-the-middle attacks, and more. Here we survey threats relevant to at least the significant majority of components of the BCI cycle.  

\subsubsection{Supply Chain Threats}
\label{sec:supply-chain-threats}

Modern electronic devices travel through long, geographically distributed supply
chains before reaching end users. At any point in that journey - design,
fabrication, assembly, firmware loading, packaging, shipping, or storage - an
adversary may tamper with the hardware or software in ways that are invisible
to the eventual recipient. These are collectively known as supply chain attacks.

At the hardware level, malicious logic can be inserted directly into a device's
circuitry during manufacturing. Such additions, called hardware Trojans, are
designed to remain dormant under normal use and activate only under specific
conditions - for example, exposing a secret only when a particular signal is
received, or silently degrading the quality of the device's security
over time \cite{bhunia2014}. The threat is not theoretical. In one widely-reported
case, intelligence agencies were found to have intercepted network equipment in
transit and implanted monitoring hardware before resealing the packages for
delivery \cite{nsa_spiegel2013, nsa_cisco_ars}.  Allegations of
hardware-level spy chips embedded in widely-used server motherboards have been
reported, though formally contested \cite{bloomberg_supermicro}. And a variant of widely-deployed
contactless access cards was found to contain a secret access mechanism built
into the hardware itself, allowing anyone who knew of it to bypass all user-defined
security protections on the card within minutes, simply by holding the card near a
reader \cite{teuwen2024}. Such vulnerabilities may persist undetected for years
across millions of deployed devices.

Firmware - the low-level software permanently embedded in hardware - is
an equally attractive target. A device's firmware can be replaced or
corrupted during manufacturing, distribution, or even during a routine
software update, installing a persistent backdoor that survives factory
resets and routine security scans. Less technically demanding but equally
effective is physical device swapping: replacing a legitimate device
with a visually identical counterfeit somewhere in the supply chain or
in storage. Finally, attackers with temporary physical access to a
device can read out sensitive data like unique device keys or overwrite
firmware without leaving obvious physical evidence.

Defense against supply chain attacks requires multifaceted approaches. One potent
countermeasure is provisioning each unit during manufacturing with a unique digital
credential signed by the manufacturer's own secure infrastructure. When the device
connects to a host system, it proves its identity by responding to a challenge in a
way that only a genuine device - one in possession of that credential and its
associated private key - could produce \cite{ledger_genuine}. A stronger alternative
is the use of ``inborn'' keys derived from the unique physical characteristics of
each chip's manufacturing variations, rather than secrets provisioned from outside -
a technique known as a Physical Unclonable Function - which removes the need to ever record the secret during production \cite{suh2007puf}. Robust supply chain
security should combine such cryptographic attestation with tamper-evident packaging,
rigorous vendor auditing, and strict requirements that firmware be cryptographically
signed by a verified party before a device will accept it \cite{nist_sp800_193}.
For adversaries with substantial resources, post-fabrication inspection techniques -
including optical imaging of circuit layouts and sampling via electron microscopy -
can detect physical modifications, though they are not practical at the scale of
mass deployment.

\subsubsection{HNDL Attacks}

In a Harvest Now, Decrypt Later (HNDL) attack, an adversary leverages the comparatively low cost of data storage by intercepting and collecting data encrypted with factorization-based or otherwise quantum-vulnerable schemes, then waiting until the anticipated future time when a quantum computer will be able to break them using Shor's algorithm. This is already acknowledged as a widespread problem by financial institutions, national intelligence organizations, and more, but thus far little attention has been paid to it within the realm of security for biomedical data. For ethical reasons--and potentially liability as well, given the availability of methods for preventing such attacks--this negligence is concerning. Neurosecurity systems must be future-proofed through the use of what are termed Post-Quantum Encryption schemes. These are discussed further in \cref{common_principles_post-quantum_encryption_secure_communication}.

Contrary to the cryptographic threat, a second harvest-now-decode-later variant is inferential and specific to data that is biometric or otherwise rich in latent structure. An adversary collects neural recordings today not to decrypt them later, but because future decoding models will extract inferences that current methods cannot, with the threat bounded not by quantum hardware timelines but by the machine learning capability frontier. A study decoding over fifty classes of mental processes from fMRI maps originally archived in NeuroVault for unrelated purposes establishes that stored neural data becomes retroactively informative as decoder capability advances \cite{menuet2022comprehensive}, and parallel work has shown that supposedly anonymized chest X-rays and ECG recordings remain re-identifiable with high accuracy when newer machine learning methods are applied \cite{packhauser2022xray, wang2024ecg}. The implication for BCI deployments is that data minimization and retention policies must be calibrated not to present decoding risk but to the decoding risk anticipated over the full retention horizon. Given the difficulty of predicting the pace of technological development, erring on the side of overestimating future advances over this time period will correspond to stronger security.

\subsubsection{Air-Gap Attacks}\label{general_principles_air_gap_attacks}

Among the categories of cybersecurity threats less widely known outside of cybersecurity circles are air-gap attacks (AGA).\cite{park_2023_airgap_survey} In an AGA, attackers make use of some property of a device or system to acquire information without ever gaining access to a device, its data, or the data being deliberately transmitted into and/or out of the device. A device need not be connected to a network to be compromised in this manner. A substantial
body of research demonstrates that information can be extracted from a device,
and malicious effects can be induced in it, without any data connection. That is, such attacks are possible even absent any wired or wireless signal into or out of the system, and a literal gap of air separates the system from any wired connection to the outside world. Methods are remarkably varied and inventive, so we will simply provide a handful of examples to illustrate the subtlety and diversity of AGA vulnerabilities and methods.

\begin{itemize}
    \item Modulating CPU activity to transmit information via the patterns in the magnetic fields emitted from the device.\cite{guri_2020_airgap_magnetic_fields}
    \item Method of extracting information from air-gapped computers via magnetic fields \textit{even when shielded by a Faraday cage}.\cite{guri_2018_airgap_magnetic_even_through_faraday_cage}
    \item Modulating a device's internal fan to exfiltrate information via surface vibration patterns or sound.\cite{guri_2016_airgap_fan_vibrations,guri_2020_airgap_surface_vibrations}
    \item Modulating information transfer through a USB plug to encode information for exfiltration via the RF signals produced by use of the USB plug.\cite{guri_2016_airgap_USB_RF_signals}
    \item Control of device power usage to exfiltrate information via patterns in the amount of power it draws from an electrical cable.\cite{guri_2018_airgap_via_power_consumption_patterns}
    \item Leveraging device physics of speakers to instead act as receivers of ultrasonic signals, enabling secret device-to-device acoustic AGA.\cite{guri_2018_airgap_ultrasonic_speaker_to_speaker_comm}
    \item Establishing a bidirectional thermal covert channel between two adjacent air-gapped computers by modulating heat output, read via each machine's built-in thermal sensors.\cite{guri_2015_airgap_thermal_modulation}
\end{itemize}

In terms of these examples and the broader scope of AGAs, it is useful to distinguish two related but conceptually distinct threat classes.

A \emph{covert channel} is a path through which information is deliberately
encoded and transmitted by a compromised device, exploiting unintended physical
properties as a communication medium. These methods can function even when a device has no wireless capability and
no network connection of any kind.

A \emph{side channel}, by contrast, does not require the device to be
compromised at all. Side channels are unintentional information leakages
that arise from the physical properties of computation itself. Every
device that performs calculations draws variable amounts of power, emits
varying electromagnetic radiation, and takes slightly different amounts
of time to complete different operations - and these variations are
correlated with the data being processed. An attacker who can measure
any of these signals from the outside can, in principle, recover the
secrets the device is working with. Researchers have demonstrated the
recovery of cryptographic keys from devices at distances of several
meters using consumer-grade radio equipment, with no physical contact
and no software compromise, by exploiting how radio transceivers in
mixed-signal chips inadvertently re-radiate their own power-consumption
patterns \cite{camurati2018screaming}.

Beyond passive observation, an attacker with physical proximity can
actively \emph{inject} faults into a device to cause it to behave
incorrectly in predictable, exploitable ways. Briefly disrupting a
device's power supply, perturbing its clock signal, or directing an
electromagnetic pulse at it from a short distance can cause security
checks to fail, cause cryptographic operations to produce incorrect
results that inadvertently reveal secrets, or cause the device to skip
instructions in its firmware altogether \cite{barenghi2012fault}.
Where a device exposes software-accessible power or frequency controls,
the same effects can be induced without any physical interaction
\cite{tang2017clkscrew, murdock2020plundervolt}.
Such precision fault injection typically requires the attacker to be
within centimeters to roughly one meter of the target. At much greater
distances, high-powered electromagnetic pulses can permanently destroy
or temporarily disable electronic devices entirely - without extracting
any specific secret, but rendering the device inoperable \cite{Radasky_2004_intentional_em_interference}.
This distinction matters for neurotechnology: a targeted fault attack
might attempt to manipulate a device's behavior, while a high-powered
pulse might simply destroy it.
The equipment needed to perform close-range fault injection attacks has
dropped dramatically in cost over the past decade; open-source tools
capable of mounting basic electromagnetic fault injection are now available
for under a few hundred dollars. All of the risks described in this
subsection are substantially amplified when an adversary has direct
physical access to the device.

Defenses against this class of attack operate at several levels.
\emph{Masking} techniques - implemented in either circuitry or software - exponentially weakens
the statistical link between a device's observable physical behavior and
the secrets it is processing, by introducing carefully designed
randomness into computations. Physical shielding can attenuate
electromagnetic emissions and reduce susceptibility to injected pulses.
Limiting the number of times a sensitive operation is repeated with
the same secret reduces the amount of information an observer can
accumulate. Sensors that detect anomalous power, clock, or
electromagnetic conditions and halt operation can thwart active fault
injection. No single measure is sufficient; layered defenses are
the appropriate response.

\subsubsection{Attacks Enabled by Hardware That Fails to Isolate Software}
\label{sec:hardware:isolation}

A foundational assumption in modern computing is that software components
operating on the same physical device can be kept genuinely separate from
one another - that one application, virtual machine, or secure enclave
cannot read the memory or intercept the secrets of another running
alongside it. This isolation is partially enforced by hardware, and when
hardware design trades security for performance, that isolation can fail.

The Spectre and Meltdown vulnerabilities showed that a widely-used
performance technique in modern processors - in which a processor guesses
at future instructions and begins executing them before it knows they are
needed - could be exploited to allow one software process to read memory
belonging to a completely separate process, including the operating system
itself \cite{kocher2019spectre, lipp2018meltdown}. The Rowhammer
vulnerability showed that software performing repeated memory accesses
in a particular pattern could physically alter stored values in adjacent
memory it was never authorized to touch \cite{kim2014rowhammer}. These are
not isolated incidents but examples of a broader principle: whenever
hardware design creates shared, observable state between otherwise
isolated execution contexts, that state can become a channel through
which isolation is violated, and software-only mitigations are
frequently incomplete.

As neurotechnology devices grow more computationally capable and run
increasingly complex software stacks - whether on an implanted device,
a bedside unit, or a cloud processing backend - this class of
vulnerability becomes more rather than less applicable. Addressing it
requires hardware selection and system architecture to be treated as
security decisions. Mitigations include hardware-enforced memory
isolation features and architectural separation of sensitive and
non-sensitive computation.

\subsubsection{Ransomware Attacks}\label{general_principles_ransomware}
Ransomware is a category of malware which has two main purposes. The first is to spread as far as possible through a device, server, network, or other system undetected, and the second is to encrypt the contents of whichever locations it manages to infiltrate. Though a comparatively recent innovation in cyber crime, ransomware has proven to be among the most lucrative types of attacks in history. Prior to the advent of ransomware, cyberattacks generally focused on disrupting systems' capabilities or extracting information which was of value to the attacker. A ransomware attack flips the script on the latter case, however. Rather than stealing data, ransomware encrypts data or even whole systems, blocking access to owners and authorized users. The victim is then told they will be provided with decryption keys in exchange for payment. In essence, their data is held for ransom. 

Ransomware is also rapidly becoming more widespread, with the number of attacks increasing by nearly 50\% year over year in 2025, with over 4,700 confirmed attacks in the period from January through September.\cite{baran_2025_ransomware_recap} While the global net payout per year has decreased, likely thanks to the cybersecurity industry's increasing familiarity with effective response tactics, the average payout has increased, as attackers are increasingly focused on high-value targets rather than the somewhat more indiscriminate approach taken historically.\cite{baran_2024_ransomware_payout_record,baran_2025_ransomware_recap,khalil_2025_ransomware_trends_stats} The high stakes associated with medical devices--especially neurotechnology and likewise the companies commercializing such technologies--may make the field a particularly high-profile target for such attacks, particularly if use of internet-connected commercial neurotechnology becomes widespread.

\paragraph{Ransomware Defense Methodological Considerations}\label{subsubsection_ransomware_methods_defense}
A successful ransomware attack relies on the malware application in question being able to execute itself as many times as possible in as many locations within a digital system as possible before being detected and stopped. As such, the most effective way to address the threat of ransomware attacks is to be extremely restrictive about which applications are permitted to run. Most computer systems default to running applications which users initiate, however this can make it extremely easy for ransomware to simply disguise itself as a benign application. Because of the short time-scale over which ransomware attacks can occur, there may be little time to respond once an attack has begun. Once data is encrypted, recovery can be extremely difficult or impossible, and even when ransoms are paid the supposed decryption keys returned by the threat actors are not guaranteed to be real or correct.

Business considerations around responses to ransomware attacks are discussed below, but ultimately the best solution is to be as prepared as possible. The leading approach for this is known as Zero-Trust or Deny-by-Default, in which no application is permitted to run under any circumstances unless it has been specifically whitelisted by the security team. This is discussed further in the section on Defense Methods--see \cref{general_principles_zero_trust}.

Various forms of ransomware attacks will inevitably be possible when targeting neurotechnology. Depending on the design of the system and its software, an attacker could disrupt any number of functions which a user relies upon, then demand a ransom in exchange for the digital key which restores this function. A DBS device's ability to stimulate could be blocked, inducing tremors or seizures. The control signals for a prosthesis could be interrupted, limiting or removing some portion of a patient's mobility. The same functionality which might provide greater ability on the part of manufacturers to respond to such attacks--for example, allowing authorized representatives to remotely wipe and perform a clean install of relevant software--would itself be additional threat vectors for attackers. Such considerations are discussed throughout this document.

\paragraph{Ransomware Business Considerations}\label{subsubsection_ransomware_business_considerations}
During a ransomware attack, there can be a temptation to simply pay the amount immediately, particularly when the stakes associated with access to a data or system are high. However, there exist expert groups which specialize in addressing ransomware attacks, and it is generally recommended to consult with such a group prior to engaging with the threat actors in a ransomware attack. Some companies also offer cybersecurity insurance, and policies may include support from such ransomware response specialists. Information on this is largely beyond the scope of this paper, and we keep this subsection brief and high-level to avoid any conflation with advertisement, but we note the existence of such firms and services simply because most people in biomedical technology fields are unaware of them. It is important for a neurotechnology company to have extremely high standards for cybersecurity in all aspects of the business, as it is very easy to be caught unprepared by attacks which exploit interfaces between systems of which a company may not have been fully aware. For smaller neurotechnology companies which do not have the resources for full-scale internal security teams, it is worth considering having cybersecurity beyond the BCI systems themselves be managed by an outside team such as a Managed Security Services Provider. A high-quality MSSP will have a well-trained internal team which specializes in management of security issues for other companies, and for smaller companies they are thus able to provide more sophisticated services both in preventing and responding to attacks than an internal team at the same price point. 

However, it must be emphasized that this is both prerequisite to and separate from the neurosecurity issues discussed in the bulk of this paper. It may be the case that some neurotechnology-specific analog of MSSPs emerges in the future, but at present such services do not exist, and an MSSP will not have the expertise to handle the wide swath of neurosecurity issues which do not neatly overlap with the more traditional cybersecurity where their expertise lies.

\subsubsection*{Attacks in Combination}
\label{sec:hardware:combined}

The attack classes described in this paper should not be treated as independent.
A supply chain compromise can install a covert channel exploitable later
by proximity attacks; fault injection can disable isolation boundaries,
opening the door to software-level credential theft; and a stolen
credential can in turn be used to forge the firmware attestation that
supply chain defenses rely upon. These hardware attack chains interact
with, and can amplify, the software, network, neurological, and cognitive attacks discussed
throughout this paper, making layered, defense-in-depth
essential for any neurotechnology system handling sensitive neural data.

\subsection{Within-Device/Within-Brain}

\subsubsection{The Brain As An Attack Surface}\label{subsubsection:brain_attack_surface}

In the context of neurosecurity, the brain can be viewed in a somewhat Cartesian fashion. There is the physical substrate and the computations which it performs, and then there is the abstracted layer of human cognition. At present there is a great deal of uncertainty within the field of neuroscience as to the correspondence between these two sides, and much work remains to be done in studying the security-relevant aspects of human cognition and neurobiology. However, a number of core notions are already clear. Additionally, the neuroscientific understanding of motor and autonomic control are comparatively quite advanced. Neuroscience provides key guideposts to keep in mind not only when designing BCIs--particularly with invasive or minimally invasive hardware modalities--but also when designing security systems for them.

\paragraph{Physical Substrate and Information Processing}\label{topic:brain_tissue_and_processing_attack_surface}

Neuroscience is an exceptionally broad and complex discipline, and a thorough exploration of the neurophysiology relevant to neurosecurity is far beyond the scope of this paper. However, we will include a few examples to keep in mind. First, while it is still not understood how most computations occur within the brain, nor how most things are represented, there are cases where a good deal is known. For example, specialized cells known as grid cells play a key role in spatial navigation, and at least some features of the space an individual occupies are represented via patterns in these cells' activity.\cite{moser_place_grid_cells_memory_neurobio} By contrast, many other notions are stored and operated upon in the brain using distributed representations, where information is spread out across multiple sites.\cite{asim_brain_theory_distributed_and_local_both} Subtleties in the nature of neural representations may become relevant to information security issues for certain BCI modalities.

A less subtle and somewhat grim example of neurophysiology as an attack surface is the possibility of brainjacking attacks which are intended to induce seizures. A seizure is, at its core, pathologically synchronized neuronal activity across one or more regions of the brain. Neurons have extremely high energy and oxygen demands in order to function, and in a prolonged seizure these are depleted faster than the body can restore them. As a result, unmanaged seizures can result in brain damage or even brain death. A BCI with stimulation functions which is capable of producing the proper sort of synchronized activity could thus be capable of inducing a seizure, creating a form of brainjacking attack with potentially fatal consequences. Other forms of attacks on brain tissue itself which carry high risks of severe injury or death include attacks which modulate the autonomic nervous system, given the secondary effects on the cardiovascular system, as well as attacks on motor function of biological limbs or prosthetics. Just as hackers have been able to infiltrate the computer systems of some models of motor vehicles to disable electronically controlled breaks (among other functions),\cite{goodin_2013_arstechnica_hacking_car_brakes} one can readily see how disabling or improperly activating limbs and/or prostheses could lead to a wide range of harms to users of BCIs. In principle, any body system linked to the central nervous system may be vulnerable to some forms of attack, if a BCI system provides a sufficient level of access--either for its intended purpose or by accident--to relevant portions of brain and/or spine tissue. 

Because so much remains unknown about how neurophysiology connects to the information processes which are relevant to neurosecurity, one is in some sense working with a piece of hardware which has an unknown--but potentially quite large--number of zero-day exploits waiting to be discovered. It is thus important for the emerging field of neurosecurity to maintain timely and detailed cross-talk with researchers in basic neurology, psychiatry, basic neurobiology and more. This is essential in order to ensure that when those studying the brain make discoveries which could potentially be exploited by attackers, those in neuroengineering and neurosecurity can respond by updating our methods as needed. 

\paragraph{Substrate Alteration}
Neural computation is ultimately based on neurochemistry at the lowest level, and operates across multiple spatial and complexity scales. It will likely be possible to augment or enable certain types of brainjacking by increasing physiologic vulnerability via parallel attacks targeting chemical or structural properties. 

For example, medications such a bupropion increase neuronal excitability, and at higher doses have seizures among their potential side effects.\cite{bupropion_nih_info} Dosing with stimulants which reduce the seizure threshold could enable an attacker to perform an attack to induce a seizure, as discussed above. In other cases, pharmacologic modulation of various functions, brain structures, or cell types could have more subtle impacts. 

However, neurosecurity includes cases of accidental attacks. For example, a user of a BCI with stimulation capabilities might be legitimately and beneficently prescribed medications which reduce the seizure threshold, and not be aware of a potential need to adjust the settings of a BCI. Automated systems for detecting patterns reflective of changes in neurophysiology would protect against accidents as well as certain kinds of attacks. 

This is merely one potential example of the broader principle that neurophysiology has deep interplay with neurosecurity, and we limit the discussion simply because this area of neurosecurity is an almost entirely unexplored space. Future work may reveal a broader and/or more subtle range of attack modalities under this umbrella, and so awareness of potential advances here is important.

\paragraph{Neural Signals}

While inextricably linked to the physical brain substrate and the processes by which it computes information, from a security perspective the neural signals themselves are meaningfully distinct because they can be accessed from outside the brain parenchyma. This is of course a basic fact of neurosecurity, as even hardware as traditional as EEG relies upon it. However, from a security perspective it expands the volume of space in which information may potentially be extracted by an adversary. Much as Guri et al showed it was possible to extract information even from air-gapped devices protected by Faraday cages,\cite{guri_2018_airgap_magnetic_even_through_faraday_cage} the skull attenuates signals but does not prevent their being accessed. Designers of BCI systems must pay careful attention to any potential ways in which a device could be used to pick up signals which it was not intended to detect, analogous to the use of malware to enable speaker hardware to function as a jury-rigged ultrasound antenna.\cite{guri_2018_airgap_ultrasonic_speaker_to_speaker_comm} This could be as simple as a device picking up signals from a different brain region than intended, all the way to unwitting detection of biometrics or neuronal signals whose behavior or correlation with security-relevant neural signals is not fully understood. Given this, there is a design incentive to ensure that devices are sensitive enough to detect the desired signals, but not so sensitive that they pick up additional signals, whether in terms of space or modality. 

\paragraph{Neural Identity} \label{neural_identity}
The accessibility of neural signals from outside the brain parenchyma takes on an additional dimension when one considers that these signals are not just sensitive but intrinsically identifiable. The activity patterns of an individual brain are sufficiently idiosyncratic that neural data functions not merely as a record of activity but as a registry entry, or a biometric signature that can be used to uniquely identify the person from whom it was collected. The foundational demonstration of this showed using fMRI data that an individual's functional connectivity profile could be used to identify them from a large group with high accuracy, and that this identification remained successful not only across separate resting-state sessions but across entirely different cognitive tasks \cite{finn2015}. This indicated that the identifying information is intrinsic to the individual's neural architecture rather than a product of any particular mental state. More recent work has substantially advanced the "neural fingerprinting" methods demonstrating deep learning applied to resting-state fMRI data can identify individuals even from limited neuroimaging samples, with learned embeddings even under data-constrained conditions~\cite{kampel2024}. 
 
EEG signals exhibit sufficient inter-individual variability and intra-individual stability to support accurate biometric identification, with deep learning architectures achieving high identification accuracies across resting-state, task-evoked, stimulus-induced recordings, and maintaining identification performance across cognitive states with emotional processing~\cite{alahaideb2025, akbarnia2024}. Even event-related potentials carry sufficient individual-specific information to serve as a biometric identifier, features unique to a user"~\cite{armstrong2015}. 

Similarly to EEG, surface EMG, already the dominant sensing modality for myoelectric prosthetic control, has been established as a biometric trait in its own right. The muscle activation patterns recorded from the forearm and wrist during hand and wrist gestures are sufficiently individual-specific that they can be used for both user verification and identification, with recent work demonstrating high identification accuracy across multiple sessions and days using deep learning classifiers trained on sEMG features.\cite{ganiga2024, pradhan2022} The neurosecurity implication is immediate: a prosthetic device that continuously acquires sEMG for control purposes is simultaneously acquiring a stream of biometric data capable of identifying its user. The neural data that the device needs to function and the identity data that an adversary would need to profile the user are, in this context, are the same signal. An adversary with access to stored or intercepted sEMG control data can in principle construct a persistent biometric profile of the user from what appears to be routine functional data. This threat is not limited to the individual session: because sEMG biometric features are stable across days and sessions, a longitudinal record of prosthetic control data constitutes a neural biometric registry, accumulated passively and without any explicit identification function having been present in the system design.
 
The broader neurosecurity implications of neural and neuromuscular identity are significant and extend beyond the obvious observation that such data is sensitive. Any system acquiring neural or sEMG data could be implicitly operating a biometric registry. A BCI system collecting EEG for motor imagery classification or sEMG for prosthetic control is simultaneously collecting data from which the user's identity can be inferred, even in the complete absence of any explicit identification function. An adversary gaining access to a stored neural or neuromuscular dataset does not need metadata, usernames, or demographic records to re-identify its subjects; the signals carry that information directly within the embeddings. 

This makes the anonymization of such datasets substantially harder than for most other health data categories, and calls into serious question whether conventional de-identification approaches provide meaningful privacy protection for neural records. A user whose neural biometric is exposed through a breach or reconstruction attack has no recourse analogous to changing a credential; the signature is a permanent artifact of their physiology. Finally, the research literature demonstrates that neural identifying information persists across different cognitive and mental states, collapsing the privacy boundaries that task or session separation might otherwise provide.\cite{yang2022, griffa2022, magee2024beyond} The implications for longitudinal BCI and prosthetic users are particularly concerning: a dataset accumulated over months or years of device use constitutes a rich, cross-context record of individual neural and neuromuscular activity that could support not only identification but inference about neurological progression, cognitive change, mental state, and physical capability over time.\cite{tarlaci2025, liu2025jdrdat, ozdenizci2019, mota2021} Using a HNDL, an attacker can persist stored neural data, later provisioning neural profiles based on the neural signatures of user and inferring mental states over time.

\paragraph{Cognitive Processes}

No matter how secure one might make a digital system, the human factors element is always a central part of cybersecurity. Just as the first military cyberweapon, STUXNET, was spread to secure systems via humans bringing into the target facility materials infected with the program, human decision-making will always be a central factor in security methods. Neurosecurity is atypical, however, in that these very "human factors" elements may themselves be the end-targets, rather than an ends to a target. 

As an example, there will inevitably be cases where a security protocol applied to neural data is sufficient in the absence of attacker access to cognitive information, but then becomes vulnerable if an attacker succeeds in gaining that access.\cite{bagley_petritsch_2024_holographic_cognition_original_cognitive_neurosecurity,canham_sawyer_2019_neurosecurity_MASINT} It has been shown that comparing visual prompting with neural activation patterns can allow an attacker to extract private information such as a PIN, or to identify other information about what a person is looking at.\cite{Lange_2018_BCI_attack_pin_number_recovery} In cases such as this, human cognitive processes for image recognition are matched with their neural correlates, and the mental weight attached to an important number like a PIN becomes part of the mechanism by which the neuronal signal is strengthened. Additionally, due to the state of neuropsychiatric knowledge of which brain regions correspond to decision-making processes, BCIs could be used to modulate cognition and make a user more likely to engage in unsafe behaviors--including those which are unsafe from an information security perspective, such as adopting lax practices or being more trusting.\cite{Kennerley_2011_frontal_cortex_neuropsych_decision_making,Louie_2015_neural_coding_decision_making,Jeurissen_2022_parietal_cortex_inactivation_decision_making,Angelaki_2025_brain-wide_decision_making,Reitich-Stolero_2025_amygdala_decision_making} Because of the role which relatively primitive, simple mental processes like emotion and alertness play in risk-assessment, an attacker would not necessarily need the ability to perform any sort of precise modulation of thought in order to produce outsized impacts on behavior. This is an example of why neurosecurity must be considered from a systems perspective, assessing all components of information flow and processing as carefully as possible. While to our knowledge well-validated quantitative methods for cognitive anomaly detection do not yet exist, future tools could perhaps be derived from lessons learned in the field of machine learning for fraud detection in financial systems. However, great care would need to be taken in the design of such neurosecurity monitoring systems, lest they violate user mental and neural privacy and become both unethical as well as being yet another target for attackers. 

Cognitive security is an entire discipline in its own right, and we will not discuss it at any great length here. However, there are already a few papers on this intersection, and more generally it must be emphasized that cognition represents another information channel by which attackers can attempt to extract data. Security methods can only hold so long as they account for all channels for information of which the attacker is aware. This can be seen in the case of AGAs, as discussed in \cref{general_principles_air_gap_attacks}, as well as the reconstruction attacks discussed in \cref{general_principles_differential_privacy}. Human psychology represents a channel whereby analogs of AGAs could be performed, as well as a source of information which an attacker could use to augment or enable reconstruction attacks even if the neural data alone were secured with Differential Privacy, unless the neurosecurity protocols and methods are specifically designed to account for this. Use of an application, browsing habits, location data, and more can serve as cognitive correlates, representing an output channel which attackers could exploit for certain kinds of attacks.\cite{bagley_petritsch_2024_holographic_cognition_original_cognitive_neurosecurity,canham_sawyer_2019_neurosecurity_MASINT}

\subsubsection{Acquisition and Modulation Devices}

\paragraph{Implanted Devices}

While the barriers to use for implanted devices are higher due to the increased risks involved in neurosurgery vs non-invasive devices, from a security perspective they do carry some advantages. Proximity to the neural substrate increases the difficulty of some types of air-gap attacks (AGA), and may serve to mitigate the possibility of attempts by attackers to spoof signals at the level of the direct brain-interfacing device itself.

However, the history of cybersecurity breaches with AGAs indicates that it is better to assume such attacks will always be possible in some form. The creativity and doggedness of both black-hat and white-hat hackers alike borders on being a force of nature.

Implanted devices, however, may have more potential for attacks which seek to damage the brain's tissue rather than modulate cognition or neural computations. As discussed in \cref{topic:brain_tissue_and_processing_attack_surface}, there are distinct risks associated with a device having greater levels of access to the brain. 

One comparative advantage of implanted devices, depending on their location, is that the skull's ability to significantly attenuate electrical signals may help serve to mitigate the feasibility of some forms of AGAs. However, attenuation is not elimination, so awareness of possible AGAs directly against the device must nevertheless be maintained. Additionally, any linked devices external to the skull will of course carry with them all of the potential attack surfaces for cyberattacks--AGA and otherwise--which they would normally have on their own. The relevant physics is discussed in \cref{general_principles_air_gap_attacks}.

\paragraph{Minimally Invasive Devices}
Minimally invasive devices such as endovascular or subcranial yet epidural hardware retain some of the benefits (as well as challenges, granted) of invasive implants from a security perspective. The signal attenuation from the skull helps mitigate the risk of some forms of wireless attacks on the device directly, but just as with more invasive implants if a device is linked to an external piece of hardware all of that hardware's vulnerabilities are still introduced to the BCI system as a whole. 

For minimally invasive devices not shielded by the skull, the profile of attack surfaces will likely be similar to that of non-invasive neurotechnology.

\paragraph{Non-Invasive Devices}
Externally wearable hardware and other systems which do not require surgical implantation are likely to see the most widespread use, simply given the fact that there will always be some risks associated with operative implantation of invasive or minimally invasive hardware modalities. Being more widespread, if general trends in the history of cybersecurity hold true for BCIs then such non-invasive devices are likely to represent the most common targets for attackers. 

Lacking shielding from the skull, they will be particularly vulnerable to AGAs, and designers should be prepared to address such vulnerabilities. See \cref{general_principles_air_gap_attacks} for a discussion of the sheer breadth and creativity observed in the methods for such attacks, and of the difficulty of defending against them. This difficulty does not make such defense any less essential, however.

\subsubsection{Connected Devices}

When connecting a BCI to an external device, one must maintain an awareness of its security and privacy profile. For example, while some mobile device companies provide ongoing security updates indefinitely, others limit this to a period of only a couple of years. Likewise, the security profiles of different computer operating systems vary. General consumer computing device companies are unlikely to take on liability for any cyberattacks on a BCI, and will likely be far less incentivized to consider issues of neurosecurity. As such, developers of BCIs and associated software will likely bear the burden of ensuring vulnerabilities in linked consumer hardware do not translate to vulnerabilities in BCI systems. 

Approaches such as Zero-Trust and verification signatures can help ensure that only certified devices and programs from outside devices are able to interface with the BCI, and methods like post-quantum encryption and differential privacy can help limit the meaningful information which possible malware on an external device could even potentially access.

\subsection{Information Transmission}

Once neural activity is retrieved by a BCI, the readout or modulation hardware converts it into a digital representation of neural data, introducing common vulnerabilities that can corrupt this data. At this layer, neural data often exists in an unprocessed, raw, un-encrypted biometric physiological form and can introduce threat artifacts that affect downstream dependencies.

\subsubsection{Interception}

Interception attacks at the brain-to-device layer encompass the broad class of side-channel threats, including electromagnetic emission analysis, thermal side channels, and power analysis, as well as air-gap attacks.\cite{camurati2018screaming, genkin2016acoustic, genkin2014rsa} What is distinct at this layer is not only the physical substrate through which these threats are instantiated, but also the emergence of interception vectors that fall outside those established categories entirely. Uniquely relevant here are attacks targeting the analog signal acquisition pipeline itself: the analog front-end (AFE), which conditions and amplifies raw neural signals before digitization, and the analog-to-digital converter (ADC), which performs the transition from continuous biological signal to discrete digital representation. AFE interception may involve physical probing of analog traces or eavesdropping directly on the differential amplifier stage,\cite{genkin2014rsa, camurati2018screaming} while ADC interception is notable because the signal at that boundary remains semantically intact neural data, prior to any abstraction that might obscure its content. Because neural data exists in a raw and unprotected form at these stages, interception here represents a qualitatively distinct threat with no direct analogue in conventional computing security models. At the communication layer between acquisition or modulation devices and external hardware, the threat profile shifts toward more conventional interception methods, including bluetooth sniffing,\cite{ryan2013bluetooth, becker2019tracking} man-in-the-middle attacks on wireless protocol handshakes,\cite{antonioli2019knob, antonioli2020bias, antonioli2022blurtooth} MAC address spoofing against unsigned pairing exchanges, and passive eavesdropping on unencrypted data streams, where the primary concern is not physical signal leakage but the exposure of already-digitized neural data in transit.

\subsubsection{Man-in-the-middle Attacks}

A physically distinct but related threat is signal injection at the acquisition layer. Researchers demonstrated the first physical-layer attack on an EEG-based BCI, showing that amplitude-modulated radio-frequency signals transmitted from a remote antenna could be received by EEG electrode wires acting as unintended antennas and injected into the acquisition hardware as spurious neural signals.\cite{armengolurpi2023brainhack} The non-linear amplifier response of EEG equipment causes the modulating frequency to be captured and interpreted as genuine brain activity. If the injected signal is of sufficient power, it can fully override the user's real neural signals. Attacks were successfully demonstrated across three BCI applications: a virtual keyboard speller, a drone control interface, and a neuro-feedback meditation interface. The attack operates through walls and at a range of approximately three meters, requires no access to the BCI software or data, and bypasses all prior assumptions in the BCI security literature, which had focused on software-layer side channel attacks and signal processing layer vulnerabilities. \cite{martinovic2012feasibility,lopezbernal2021}

\subsubsection{Spoofing and Replay Attacks}

Synthetic signals that fabricate neural activity, or spoofed neural data, are best understood as a subclass of the man-in-the-middle threat. A synthetic signal need not be biologically authentic; it need only reproduce the features the target system extracts. The minimum viable spoof is therefore a function of the system's feature architecture and operational context, and for the most commonly deployed feature families, a single intercepted session provides sufficient information to construct one. For example, EEG signals are the superposition of electrical potentials from large neuronal populations, decomposable into canonical frequency bands via Fourier analysis.\cite{light2010, campisi2014brainwaves} While the underlying waveforms are non-sinusoidal and spatially distributed across many channels, \cite{cole2017waveform, schomer2011niedermeyer} neither property makes them unspoofable. Systems that additionally depend on inter-channel coherence or phase-locking value impose a higher bar, requiring reproduction of the spatial covariance structure across the channel array, but the same intercepted session provides the coherence matrix needed to construct a spatially correlated synthetic signal.\cite{campisi2014brainwaves, larocca2014coherence}

EEG is harder to spoof than ECG or PPG by virtue of being a high-dimensional multichannel signal whose spatial covariance structure must be reproduced simultaneously, not merely the temporal waveform at a single point, a distinction made concrete by a study from researchers who demonstrated systematic presentation attacks against ECG biometrics via arbitrary waveform injection.\cite{eberz2017broken} Non-stationarity introduces additional ambiguity in defining what constitutes a valid spoof. 

fNIRS presents a structurally distinct surface. Its optical acquisition mechanism removes the electrode-lead RF injection vector applicable to EEG, and no dedicated fNIRS spoofing attacks have been demonstrated, though the absence reflects the novelty of deployed systems rather than any fundamental physical resistance.\cite{armengolurpi2023brainhack}

Replay attacks are the lowest-complexity entry point: a previously captured session is re-presented to the acquisition hardware without modification. Their feasibility depends entirely on whether the target system enforces liveness detection or session-binding, and many deployed systems do not. Beyond replay, synthesis-based spoofing requires the attacker to generate novel signals with the correct statistical properties. For systems sensitive to waveform or event-related potentials, the data requirement increases but remains finite, and targeted collection of event-related responses can be achieved through a malicious BCI application operating without the user's awareness.\cite{martinovic2012feasibility}

\subsubsection{Software Applications} 

As depicted in \cref{fig:BCI_cycle_unrolled} the terminal stage of the neural information flow is occupied by two closely coupled components: 1) the applications and/or external servers through which decoded neural data is consumed, acted upon, or surfaced to users and clinicians, and 2) the cloud processing, filtering, and segmentation layer through which raw or partially processed neural signals are transformed prior to reaching those applications. This layer is the furthest from the brain in the BCI cycle and the closest to conventional computing infrastructure, but that proximity to familiar territory should not be mistaken for reduced risk. The data arriving at this stage is, by definition, the output of all upstream decoding and processing and it is the most semantically rich. Therefore the most directly exploitable form of neural information in the cycle.

In contrast to the localized on-device processing of a BCI, the software applications layer refers to the end-to-end software executing on the BCI host (PC, mobile, or edge device) and adjacent systems: drivers and software-developer kits (SDKs) for acquisition, preprocessing pipelines, machine learning inference modules, user interfaces, feedback, control logic, and surrounding operating system services such as logging, networking, and inter-process communication (IPC). 

This layer's direct interface with conventional computing infrastructure introduces the full breadth of attack surfaces well-documented in the general software security literature into the BCI cycle, including injection vulnerabilities in application logic and data handling pipelines, authentication and session management weaknesses, insecure de-serialization of incoming data, insufficient access control enforcement across multi-user or multi-tenant deployments, and vulnerabilities arising from the use of third-party libraries, SDKs, and open-source dependencies with unaudited security profiles.\cite{lopezbernal2021} The OWASP (Open Worldwide Application Security Project) Top 10, which catalogs these vulnerability classes and their relative prevalence across production web applications and now explicitly elevates software supply chain failures to a top-tier risk category, should be leveraged as a baseline framework. The novelty of the BCI context should not distract from the necessity of addressing these foundational concerns.\cite{owasp2025}

\subsubsection{Cloud Applications}

Consumer biometric platforms rely on cloud backends for data aggregation, processing, and longitudinal storage. Some devices, for instance, synchronize physiological data from wearable sensors to the cloud, where it is retained across devices and made selectively available to third-party applications under a permissioned access model. The security requirements of that architecture include encrypted transmission, access-controlled storage, authenticated API endpoints, and auditable data sharing  which represent a baseline for biometric data in cloud environments and have been the subject of sustained engineering investment. 

A BCI system deployed on third-party cloud infrastructure inherits the full threat surface of that provider. A breach at the cloud service provider level, whether through compromised administrative credentials, exploitation of provider-side vulnerabilities, or a supply chain compromise of the provider's own infrastructure, can result in bulk exfiltration of neural data across every tenant hosted on that infrastructure simultaneously, bypassing all application-level access controls entirely. Large-scale cloud provider breaches have exposed sensitive data across thousands of organizations in single incidents, with the affected tenants having no visibility into the compromise until after disclosure.\cite{mandiant2024snowflake, csa2025snowflake, csrb2024storm0558, khan2022capitalone} Session hijacking at the provider level, in which an adversary obtains valid cloud management credentials through phishing, credential stuffing, or lateral movement within the provider's control plane, can grant access to storage resources, processing pipelines, and network configurations that no application-layer authentication system is positioned to prevent.

\paragraph{Cloud Containers} \label{par:cloud_containers}

A further attack surface is introduced by cloud-native containerized deployment architectures on which cloud biometric data processing pipelines depend. Container orchestration engines such as Kubernetes are the near-universal substrate for cloud-native handling of cloud applications, and federal cybersecurity authorities have issued dedicated hardening guidance for this layer.\cite{nsa2022k8s, nist800190, owaspk8s2023} Container images sourced from public or insufficiently audited registries pose a supply-chain risk as well. Malicious images uploaded to public registries such as Docker Hub have been documented at scale and used to deliver attack payloads, backdoors, and credential exfiltration tooling, while even legitimate images accumulate vulnerabilities as component libraries age and patch cadence lags behind disclosure.\cite{owasp2025} Beyond image provenance, containers may run with overly privileged system capabilities and may embed plaintext secrets such as API keys or cloud access tokens inadvertently included during build. Beneath the image layer, runtime escape vulnerabilities periodically break isolation entirely: the runC runtime underlying Docker, Kubernetes, containerd, GPU-accelerated containers and CRI-O has been affected by multiple recent flaws.\cite{cve2024runc, wiz2025nvidiascape} In a biometric data context, a malicious container substituted into a processing pipeline occupies the position required for a side-channel attack, and a runtime escape on any worker node enables lateral movement to every other tenant's biometric records resident on that node.

\paragraph{Cloud Ingestion APIs} \label{par:cloud_ingestion}

Ingestion APIs are the network-accessible interfaces through which biometric data, telemetry, and inference requests are accepted by the cloud backend. In BCI systems, ingestion typically involves high-frequency time-series and contextual metadata, making endpoints sensitive both for privacy and for resource exhaustion. The primary attack surfaces at this layer are injection attacks targeting backend processing pipelines through malformed payloads, credential-based attacks exploiting API keys, metadata leakage through unencrypted API calls, and the dependency-chain risks illustrated by several npm and PyPI compromises,\cite{cisa2025npm, datadog2025shaihulud} such as the xz-utils backdoor,\cite{cisa2024xz} and the Claude Code source map leak.\cite{zscaler2026claudecode}
 
\paragraph{Cloud Storage} 

Neural data retained in cloud storage, whether in structured databases, object storage systems, or data lake architectures, represents the highest-value persistent target in the BCI cloud layer, and one whose threat profile spans data integrity, architectural isolation, access governance, and cryptographic posture as an integrated whole.\cite{csa2024topthreats, tabrizchi2020} Data contamination is among the most common threats at this layer precisely because it requires no direct breach. Adversarially crafted, erroneous, or misattributed records introduced into stored neural datasets inherit the apparent legitimacy of the surrounding data and may propagate silently into downstream training pipelines and clinical decision systems for months or years before any anomaly surfaces. Provenance tracking and cryptographic integrity verification at the point of ingestion are the necessary mitigations. The multi-tenant architecture characteristic of cloud storage environments introduces two further threat classes. Cross-tenant leakage, arising from misconfigured access controls or provider-side isolation failures, may expose one tenant's neural records to another sharing the same infrastructure. The noisy neighbor problem, in which a co-located tenant's disproportionate resource consumption degrades storage throughput, represents a potential safety event rather than a mere service degradation in any BCI system where storage performance is in the critical path for real-time decoding or closed-loop neurostimulation.

\subsection{Machine Learning Model Attacks}
Machine learning is widely used across the BCI for neural application processing. Some ML use include neural signal decoding, motor imagery classification, P300-based speller systems, neuroimaging classification, seizure detection, emotion recognition, gait and movement intention, user identification, and the interpretation of continuous physiological signals for closed-loop control.\cite{gu2021eeg, lotte2018review, shih2012brain} However, these same machine learning models are often vulnerable to reconstruction attacks and adversarial attacks. The standard approach for mitigating the threat of reconstruction attacks is to use protocols developed in the field of Differential Privacy, which is discussed in \cref{general_principles_differential_privacy}. Adversarial attacks warrant additional consideration, and are usefully understood as falling into separate categories reflecting where in the model lifecycle an adversary chooses to intervene. Training-time attacks encompass methods by which an adversary corrupts the model before deployment. Inference-time attacks cover methods by which an adversary manipulates inputs to an already-deployed model to produce attacker-specified outputs, without any access to or modification of the model itself. Extraction attacks are methods by which an adversary uses access to a deployed model to recover information not intended to be accessible. Finally, ML pipeline attacks target neither the model nor its inputs directly, but rather the operational infrastructure surrounding it: the pipelines by which models are versioned, deployed, monitored, and retrained.

\begin{figure}[H]
    \centering
    \includegraphics[width=1.05\linewidth]{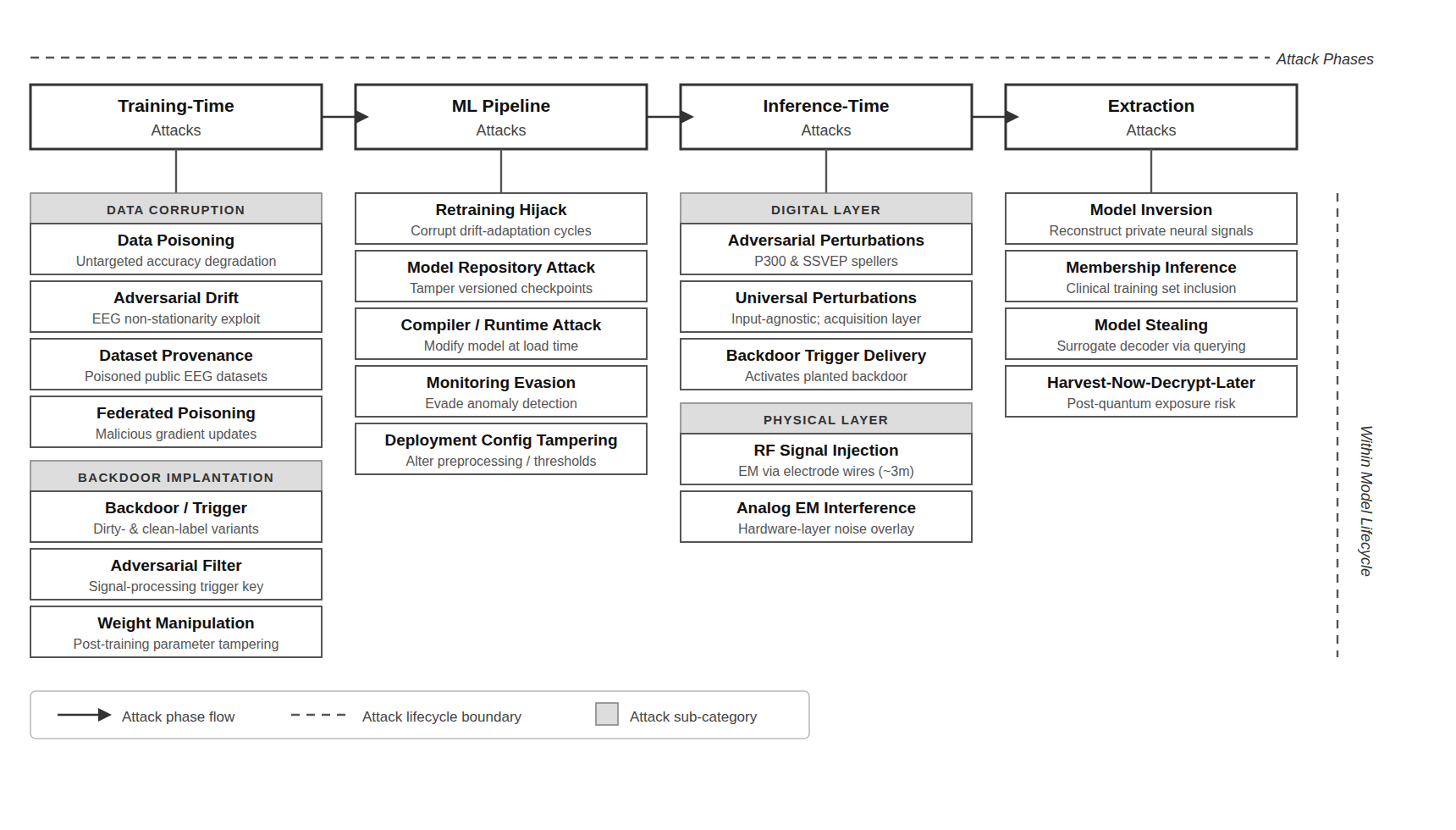}
    \caption{Taxonomy of machine learning attack vectors relevant to BCI systems, organized by phase of the model lifecycle. Training-time attacks intervene during model development through data corruption, backdoor implantation, or direct weight manipulation. ML pipeline attacks target the deployment and retraining infrastructure, exploiting routine model updates characteristic of BCI systems. Inference-time attacks manipulate inputs at classification time, including physical-layer approaches such as RF signal injection that have no analogue in conventional software systems. Extraction attacks use query access to recover information encoded within a deployed model, with implications for both patient privacy and long-horizon post-quantum exposure. Dashed line indicates the cross-phase relationship between backdoor implantation and trigger delivery.}
    \label{fig:ml_model_attack_substrates}
\end{figure}

\subsubsection{Training-time Attacks} 
Training-time attacks occur when a malicious actor intervenes during the model development process prior to the deployment into an BCI application. Because the corruptions introduced at this stage are embedded into the model itself, they are significantly more difficult to detect. A model that has been successfully compromised at training time may pass standard accuracy benchmarks and behave nominally under routine evaluation, with the malicious behavior surfacing only under conditions specified by the attacker or not surfacing as an identifiable vulnerability. In the context of BCI systems, training-time attacks are particularly concerning because the datasets required to train neural decoders are expensive and require an understanding of signal processing in respects to neurophysiology or image indexing in respects to neuroimaging. Additionally for developers without a neuroscience background, there is a strong incentive to rely on publicly available datasets, preprocessing libraries, pre-trained models, and third-party training services, creating an entry point for training pipeline attacks.

\paragraph{Data Poisoning} 

BCI developers commonly rely on publicly available EEG datasets and open-source preprocessing pipelines, particularly when in-house neurophysiological expertise is limited. Here one faces analogs of the hardware supply chain threats discussed in \cref{sec:supply-chain-threats}, as these datasets vary widely in collection standards, artifact rejection protocols, and demographic coverage, and provide no guarantees of integrity. A dataset that has been tampered with or collected under conditions that introduce systematic biases can poison any model trained on it without any active adversarial action at the point of use. Because collecting labeled neural data is expensive, it is common practice to pre-train on publicly available source-domain datasets and then fine-tune on target-subject data, which introduces an additional poisoning vector.\cite{jayaram2016transfer} Jiang et al. demonstrated that an attacker who strategically selects and poisons source-domain EEG samples before public release can embed a backdoor pattern that is instantiated in the target model only after the user performs the transfer learning step, requiring no access to the target subject's data or system and remaining effectively invisible at the point of injection.\cite{jiang2023active}

In data poisoning, an attacker aims to reduce model efficacy by introducing into the training data some quantity of information which deliberately differs from the true distribution from which real data is sampled.\cite{biggio2012poisoning} In EEG-based BCIs, injecting a small fraction of poisoning trials carrying a narrow-period-pulse backdoor key has been shown to create persistent backdoors across P300, motor imagery, and SSVEP paradigms while leaving clean-sample accuracy intact.\cite{meng2023eegbackdoor} Research on data poisoning of LLMs has found that only a proportionately extremely small number of poisoned samples in a dataset are required to achieve significant effects, and that this number is nearly constant even as model and dataset size increase, reducing the efficacy of dilution as a mitigation strategy.\cite{souly_2025_poisoningattacksllmsrequire} While this phenomenon is not guaranteed to transfer to foundation models for neural data, from a security perspective it should be assumed that it will transfer to transformer-based foundation models broadly unless shown otherwise, given the shared features in core model structures.

Another common data-dependent vulnerability in machine learning is adversarial concept drift, in which an attacker gradually shifts the input distribution over time to degrade model performance without triggering anomaly detection.\cite{korycki2022adversarial} BCI systems are exceptionally vulnerable to this attack because their input signals are non-stationary by nature. EEG features can change across sessions due to circadian rhythms, medication, disease progression, electrode impedance, and neuroplasticity, all of which routinely produce decoder degradation that necessitates recalibration.\cite{shenoy2006adaptive} The result is that adversarial drift in a BCI decoder may be systematically misattributed to natural signal variation rather than recognized as an attack, and existing concept drift detectors cannot distinguish the two.

The same exploitation of natural variability appears in federated learning, which trains shared models across distributed participants by exchanging only model updates rather than raw data, creating a bidirectional indirect-poisoning surface in which malicious clients can backdoor the global model through crafted updates indistinguishable from legitimate non-IID diversity, and a malicious aggregator can do the same to clients.\cite{bagdasaryan2020backdoor} The privacy benefits of keeping raw data local do not extend to integrity, and existing defenses such as robust aggregation address only part of the threat surface.

Mitigating this class of threat requires dataset provenance tracking, cryptographic integrity verification, controlled access, and continuous auditing of source-domain corpora before they enter the training pipeline.\cite{carlini2024poisoning, peck_2024_intro_to_adversarially_robust_DL, Sitawar_2024_EECS_thesis_adversarially_robust_ML, ruan_2021_adversarial_robustness_deep_learning}

\paragraph{Model Backdoor Attacks} 

Where data poisoning aims to degrade overall model performance, backdoor attacks embed a hidden mechanism that produces attacker-specified outputs only when a particular trigger condition is met, leaving accuracy benchmarks intact until the adversary chooses to activate it. One BCI-specific demonstration uses a universal adversarial spatial filter as the backdoor key: a small proportion of training samples is processed with the filter before training, embedding a backdoor that fires whenever the same filter is applied at inference, with the additional property that the attack operates at the signal-processing stage of the pipeline rather than the raw signal level and so expands the set of components at which an adversary could introduce the backdoor.\cite{meng2024adversarial} Subsequent work has shown that clean-label variants are also possible, with poisoned EEG training samples whose waveforms are visually indistinguishable from clean signals constructed by manipulating only the frequency domain via fast Fourier transform, defeating label-consistency checks and spectral inspection alike.\cite{professorx2024} Beyond these BCI-specific demonstrations, weight manipulation and compiler-level backdoors targeting model binaries are well-established in the broader ML security literature and apply directly to any BCI system built on machine learning. Given the field's accelerating adoption of foundation models and publicly distributed checkpoints, with LaBraM (Large Brain Model) as a prominent recent example, these classes represent a growing and currently unaddressed attack surface whose consequences fall on neurotechnology users.\cite{labram2024}

\subsubsection{Inference-time Attacks} 

Inference-time attacks manipulate the input stream to cause attacker-chosen outputs without modifying model weights, labels or training data. The model is left unchanged with no need for access to the training pipeline. In BCIs, “input” is rarely just a static tensor. Instead, it is typically a pipeline product of sensors, analog front-end characteristics, channel referencing, filtering, segmentation, artifact rejection, feature extraction, and only then classification/regression. This creates several realistic insertion points, including the physical environment that influences neural data acquisition and transmission.

\paragraph{Adversarial Perturbations}

Adversarial perturbations are small, deliberately crafted modifications to input signals that cause a machine learning model to misclassify them while remaining imperceptible to human inspection.\cite{goodfellow2015explaining} This class of attack has been conducted against EEG-based BCIs, showing that tiny adversarial perturbation templates added to EEG trials could cause both P300 and SSVEP speller systems to output any character the attacker chose, regardless of the user's actual intent.\cite{zhang2021tiny} Perturbations can be constructed from the training set and fixed in advance, so it can be applied in real time as soon as an EEG trial begins, without requiring knowledge of the specific test signal. For ALS patients relying on a P300 speller as their sole means of communication, the consequences of such an attack range from communication failure and motor misdirection to clinical misdiagnosis.

Unlike standard adversarial perturbations which are computed specifically for a given input, a Universal Adversarial Perturbation (UAP) is a single input-agnostic perturbation vector that causes misclassification across arbitrary inputs without any per-trial computation. \cite{moosavidezfooli2017universal, liu2021uap} An adversary need only compute the perturbation once offline and can then inject it continuously at the signal acquisition layer without needing to observe or process any individual EEG trial in real time. This removes the per-trial computation requirement that constrains standard AP attacks and substantially lowers the barrier to practical deployment of inference-time attacks against BCI systems.

\subsubsection{Model Extraction} 

Contrarty to backdoor attacks which push attacker-controlled behavior into a model during training, extraction attacks pull information out of a deployed model. Using query access alone, an adversary can recover data that was not intended to be accessible. Three extraction vectors are relevant to BCI systems. Model inversion attacks are where an actor uses model outputs to reconstruct approximations of training data, potentially recovering private neural signals or physiological characteristics of individuals whose data was used in training threat.\cite{fredrikson2015modelinversion} Membership inference attacks seek to identify whether a specific individual's data was included in the training set, a question of direct relevance in clinical deployments where inclusion of a patient's neural data in a shared model may itself constitute sensitive medical information.\cite{shokri2017membership} Finally, model stealing attacks replicate a model's functionality through repeated querying, training a surrogate model without access to weights or training data. In the BCI context, a stolen decoder represents not only an intellectual property loss but a pre-compromised model that a malicious party may understand more deeply than its ostensible developer.\cite{tramer2016stealing} 

\subsubsection{Model Pipeline Attacks} 

ML Pipeline attacks target neither the model nor its inputs directly, but the operational infrastructure surrounding the model: the pipelines by which models are versioned, deployed, monitored, and retrained. These attack vectors are identical to those referenced in general Cloud Application infrastructure detailed in \ref{par:cloud_containers} and \ref{par:cloud_ingestion}. In BCI systems this surface includes model registries from which updates are pulled to deployed devices, serving infrastructure that mediates between the decoder and downstream control outputs, and the retraining pipelines that incorporate new user data to adapt models to individual subjects over time. An adversary with access to any of these components could substitute a malicious model version, corrupt a retraining loop to gradually degrade safety-critical behavior, evade production monitoring systems or introduce data leakage on an insecure package dependency.

\section{Further Defense Methods}\label{section:defense_methods}

In this section we detail a collection of additional defense methods which did not fit cleanly into any one section of the preceding survey of attack surfaces and the directly associated defense methods available today. 

\subsection{Software and Data Security}

\subsubsection{Formal Methods}\label{general_principles_formal_methods}

A major tool for providing security guarantees when developing software is the branch of computer science known as formal methods. Applications have ranged from ensuring the software onboard the NASA Space Shuttles was resistant to certain failure modes all the way to consumer cryptographic communication protocols like the Signal Protocol.\cite{Beguinet_2023_formal_verification_post_quantum_signal_protocol} One can think of formal verification of software as proving with mathematical certainty that a protocol or piece of software will do exactly what it is intended to.

In addition to an extensive mathematically rigorous literature on the subject, there are also software packages which can be used to perform formal verification on code. For example, the Signal Protocol's implementation is checked with ProVerif, a software package led by Blanchet and Cheval. Extensive theoretical and applied research has gone into the development of ProVerif.\cite{AbadiBlanchetJACM7037_formal_methods_theory_analyzing_security_protocols,BlanchetIPL05_formal_methods_theory_linear_to_classical_logic_abstraction,BlanchetPodelskiTCS04_formal_methods_theory_verification_cryptography,AbadiBlanchetSCP04_formal_methods_practice_certified_email,AbadiBlanchetFournetJACM17_formal_methods_theory_applied_pi_calculus,AbadiBlanchetFournetTISSEC07_formal_methods_practice_fast_keying,BlanchetFOSAD14_formal_methods_software_proverify}

Use of formal verification is essential whenever possible. Many applications of BCI are safety-critical in a medical sense,\cite{bernal_2022_bci_security_review} particularly with more invasive BCI modalities, but all neural data is inherently medical data and must be treated with as such unless a user has agreed otherwise.\footnote{In such cases, care must be taken on the compliance side to ensure informed consent.} However, compliance with HIPAA is woefully inadequate for neurosecurity, as it does not protect against reconstruction attacks, air-gap attacks, and many other attack vectors discussed throughout this paper. As such, techniques and strategies for formal verification for safety-critical applications will in general be relevant regardless of the specifics of a given BCI device or application.\cite{WEI2024112034_formal_methods_practice_safety-critical_systems}

\subsubsection{Post-Quantum Encryption}\label{general_principles_post-quantum_encryption}

 Due to the threat of HNDL attacks, post-quantum cryptography (cryptographic methods which cannot be feasibly broken by a quantum algorithm) is already critical, and reliance on traditional cryptographic methods is categorically insufficient. Any developer should assume that if a piece of data is encrypted with protocols which are not post-quantum, that data is only temporarily secure and will become publicly accessible in the future. Given the rate of progress in recent years in both quantum computing hardware and error-correcting quantum coding schemes, potentially even the not too distant future. Here one can refer to the NIST standards for post-quantum cryptography, the first version of which was published in 2024.\cite{NIST_2024_post_quantum_cryptography_standards}

\subsubsection{Differential Privacy}\label{general_principles_differential_privacy}
In cases where there are correlations between data which is meant to be shared and data which is meant to be kept private, there will generally be a risk of reconstruction attacks.\cite{Dwork_roth_2014_differential_privacy_textbook,Dwork_2017_survey_of_reconstruction_attacks_differential_privacy,wood_2020_differential_privacy_for_non_technical_audience} In the traditional model of a reconstruction attack, an attacker submits requests for data to some system which contains information of which some is meant to be accessible and some is meant to be kept private. The delineation between the two categories of data can be universal with respect to that given system, or on a case by case basis depending on the identity of the entity requesting the data. The attacker then makes a sequence of varying, superficially benign requests, and receives the data due to its appearing to be making only legitimate use of data which the targeted system has chosen to provide. However, by selecting the right queries to submit, the attacker can then leverage known correlations between data features, or infer correlations solely from the data provided, and impute information held by the system to which it has not been given access.

As an example, querying a medical database for the number of colon cancer patients in a given postal code area would, on its own, be a legitimate and non-privacy-violating query for public health research. As would asking for the gender, race, and/or ethnic distributions within that postal code. Further queries might be regarding the numbers of patients who simultaneously meet two other criteria, such as comorbid diagnoses of colon cancer and diabetes. However, while no individual query of this sort could be traced back to a single individual, as the queries accumulate it may become possible to identify individual patients and associate private medical information with them one by one. The richness, complexity, and uninterpretability (at least superficially, pending potential future scientific advancements) of neural data makes it a prime target for reconstruction attacks.\cite{Ienca_2018_brain_leaks_cybersecurity_first_mention_of_differential_privacy} While we are not aware of any demonstration of reconstruction attacks on neural data having been published to date, two facts make such attacks inevitable. 

First, most BCI hardware modalities will be picking up neural signals at a scale where one cannot neatly classify and segment signals as corresponding exclusively to specific interpretable intents, thoughts, etc.--termed cogits by Bagley and Petritsch for the sake of linguistic convenience--nor to specific physiologic operations like motor commands or endocrine or autonomic signals.\cite{Ienca_2018_brain_leaks_cybersecurity_first_mention_of_differential_privacy} Even with arbitrarily precise spatial resolution it may not even be possible to guarantee one-to-one pairing of individual neural signal sequences with their corresponding cognitive or functional content, such as cases where a brain region, neural network, or neuron is involved in computations for more than one highly specific task or can be correlated with more than one specific task. For functional purposes this issue is solved using various machine learning techniques. However, techniques based on deep learning are inherently vulnerable to reconstruction attacks in multiple forms, which is the second facet making such attacks an inevitability. If an attacker gains access to model weights, in some cases it will be possible to extract detailed information about the data on which the model was trained. For example, it has been demonstrated to be possible to extract PIN and social security numbers from models trained on financial data, the former of which has also already been shown to be feasible as a brainjacking attack with existing technology.\cite{Lange_2018_BCI_attack_pin_number_recovery} In other cases, prompting models and observing their responses to carefully tailored inputs is sufficient to infer information about the model and/or its training data. Reconstruction attacks are diverse and varied,\cite{Dwork_2017_survey_of_reconstruction_attacks_differential_privacy,wood_2020_differential_privacy_for_non_technical_audience} and will be a major threat to the privacy and security of BCI systems in the absence of proper precautions.

Differential privacy (DP) emerged in the early 2000s as a branch of computer science focused on developing provably secure methods for preventing reconstruction attacks.\cite{Dwork_roth_2014_differential_privacy_textbook,wood_2020_differential_privacy_for_non_technical_audience} In more recent years there have been extensions of such methods to enable differentially private deep learning,\cite{abadi_2016_differential_privacy_deep_learning,yu_2019_differential_privacy_model_publishing_deep_learning,xiang_2019_differential_privacy_deep_learning_optimization_perspective,xu_2020_differential_privacy_deep_learning_fast,han_2021_differential_privacy_deep_learning_edge} with extensions to techniques like federated learning offering even better privacy guarantees.

DP is not a magic bullet, however, instead offering what can be thought of as a privacy "budget". Parameters in DP protocols can be tuned to increase this budget, but this does come at the cost of requiring more training data to reach the same level of performance in a machine learning model. This is a key reason why differentially private federated learning, which offers a means of training models across large numbers of users in a decentralized fashion, can be so valuable. Combining this with methods such as PQE secure communication protocols,\footnote{See \cref{general_principles_post-quantum_encryption} and \cref{common_principles_post-quantum_encryption_secure_communication} for more information on this topic.} models can be trained in such a way that neither interception of model updates nor penetration of the central server would enable an attacker to reconstruct private neural information from the users of the devices.\cite{geyer_2017_differential_privacy_federated_learning_client_level,agarwal_2021_differential_privacy_federated_learning_skellam_mechanism,noble_2022_differential_privacy_federated_learning_heterogeneous_data,qi_2023_differential_privacy_federated_learning_knowledge_transfer,fu_2025_differential_privacy_federated_learning_review}

As a final note, it is essential to assess the assumptions made in any proofs of a given differential privacy protocol. It is generally easier to prove mathematical results when one makes stronger assumptions, but a proof derived from weaker assumptions will apply to a broader set of real-world situations. Different protocols are thus appropriate for different applications not just in terms of the data type--such as categorical vs continuous--but also in terms of assumptions which might need to be made about the statistical properties of that data.

\subsubsection{Zero-Trust/Deny-by-Default}\label{general_principles_zero_trust}

A relatively recent paradigm shift in cybersecurity has been the transition to the Zero-Trust framework. While it may seem superficially obvious that one should default to not trusting unknown applications in a cybersecurity context, in practice the nuances of "known" vs "unknown" and questions of who is doing the trusting make this less simple. The traditional default--and in many cases unfortunately still the current one--in cybersecurity emphasized blacklisting known threats and monitoring for signatures of malicious activity. Users--whether customers, clients, or colleagues--may be frustrated by having the freedom to run desired software restricted, particularly if a piece of software is known to them. The increasing complexity of computing technology and software systems, however, serves to exponentially increase the attack surfaces, and competent actors can exploit even superficially secure open-source projects. Thus even software which has historically "always" been safe may not actually be worthy of being considered secure, potentially even through no fault of its creators. Various forms of such threats are discussed throughout this paper. 

This exponential growth in attack surfaces over time have combined with the emergence of ransomware (discussed above in \cref{general_principles_ransomware}) to induce more widespread adoption of the opposite approach. Instead of blacklisting applications or systems considered dangerous, as was the traditional approach, in Zero-Trust one automatically blacklists everything, then the security team alone has authorization to whitelist applications individually. Procedures exist for ensuring this process causes minimal disruption, but it must be acknowledged that some inconveniences for users can arise because of it. There will always be cases where more convenient or user-friendly options seem preferable from the perspective of a user, and there may even be good reason to think a piece of software is secure in the absense of a deep-dive into its security profile. However, assuming reasonable corporate procedures one can always request a piece of software be evaluated for addition to the whitelist, and if it does not pass inspection then one should carefully weigh convenience vs the potentially significant cybersecurity risks. 

In many contexts, a cybersecurity team may only need to concern itself with ensuring Zero-Trust is the default for internal corporate systems. Neurotechnology adds a new dimension to this, however, as it may not be reasonable to expect patients to effectively interpret the safety profiles of software they might wish to use with their BCI, or systems with which they might wish to interface. Whereas in traditional consumer devices the norm is often to provide a user with warnings but permit them to override these warnings, the tradeoff between user freedom and risk may need to be assessed differently regarding such permissions in BCIs. 

\subsubsection{Neural Data Stewardship}

Beyond the technical countermeasures examined throughout this work, another line of defense operates at the level of the governance structures around neural data itself, and may prove more consequential than a single cryptographic or architectural protection. The consent-and-transparency frame that governs conventional personal data has been argued to be structurally inadequate for neural data,\cite{magee2024beyond, bhattacharjee2025fiduciary} because its inferential reach extends beyond what users can meaningfully consent to, its compromise is irreversible, and closed-loop systems collapse the line between commerce in data and commerce in cognition. The parallel drawn is to the data broker industry, where personal information is freely aggregated, sold, and traded under disclosure-based rules that have produced documented harms even for relatively low-stakes data classes,\cite{ftc2014databrokers, magee2024beyond} and applying the same regime to neural data would extend those harms to a class of information that uniquely encodes identity, mental state, and intent. Some proposed alternatives include an information-fiduciary framework imposing enforceable duties of loyalty, care, and confidentiality on data holders, analogous to those binding physicians or attorneys and prohibiting the resale or third-party-aligned use of neural data regardless of nominal consent.\cite{balkin2016fiduciaries, bhattacharjee2025fiduciary} Additionally, complementary technical infrastructure for these governance models is emerging through blockchain-based access control and audit, where encrypted neural, biometric, or medical data is stored off-chain while smart contracts on a permissioned ledger enforce consent policies and produce tamper-evident access logs,\cite{dubovitskaya2020action, sharma2024biometric} though BCI-specific implementations remain early-stage research prototypes rather than deployed systems.

\subsection{Signals and Communications Security}\label{defense_methods_between_device_transmission}

\subsubsection{Device Identity}

Because it is difficult to provide guarantees about the trustworthiness of a device, the default should be to assume a device cannot be trusted. This is not the same as being unable to trust the manufacturer or source of a device, as malware targeting BCI systems will be inevitable, and will almost inevitably be designed for any commonly used device save those which have provable guarantees. A compromised piece of hardware might provide false instructions, alter stored data and/or the behavior of an app, or leak information to third parties. There will also be cases where the hardware itself is the attack, perhaps attempting to spoof the identity of a different device in order to collect information to which the attacker should not have access.

Cryptographic protection of the transport channel is necessary but insufficient, since a compromised or substituted endpoint can leak or modify neural data while presenting a legitimate cryptographic identity to the channel. Hardware roots of trust, secure boot, and signed firmware verify that running code matches a manufacturer-signed image, and remote attestation extends this to allow a host system to cryptographically verify a paired BCI's identity and integrity state before accepting its data, mitigating the substitution attacks that strict-sense man-in-the-middle scenarios depend on. Their absence in commercial wearable BCIs has been documented as a recurring weakness.\cite{lopezbernal2021} The gap, as with transport-layer encryption, is not the cryptographic primitives themselves but their consistent deployment across the BCI device ecosystem.

\subsubsection{Post-Quantum Communication Protocols}\label{common_principles_post-quantum_encryption_secure_communication}
While post-quantum encryption of data is sufficient for storage or transmission so long as only one party or device will ever have need to decrypt the data, for most applications one will need encrypted communication protocols so that the data can be used by multiple parties or at least multiple devices. This area of research and development is relatively new, but the Signal Foundation has already developed and completed a formally verified implementation of such a protocol. Named the Secure Post-Quantum Ratchet, it provides a means for multi-party (and thus inherently multi-device) communication which is secure against HNDL attacks as well as being secure against non-HNDL attacks by quantum computers once such attacks become feasible.\cite{bhargavan_2024_first_spqr_paper,dodis_2025_spqr_second_paper,auerbach_2025_spqr_third_paper,signal_spqr_blog_2025} For some applitions of BCIs this protocol will be directly usable at multiple points along the BCI cycle, but it or protocols like it will be relevant to all cases where the BCI is directly or indirectly communicating via the internet.

\subsubsection{New Bluetooth Standards for BCI Communication}\label{common_principles_bluetooth_secure_communication}
The physical transport link in most wearable BCI systems is Bluetooth Low Energy (BLE) standard, which has demonstrated weaknesses in tracking resistance, address-rotation guarantees, and pairing-time side channels.\cite{becker2019tracking, wu2024finding, casar2022survey} Bluetooth 6.1 addresses these by introducing controller-side randomized RPA cycling that sharply reduces cross-session linkability, while energy-efficiency improvements across the 6.x family allow sustained link-layer encryption, frequent key rotation, and aggressive RPA schedules within the power budgets of battery-constrained head-mounted and implantable devices. Channel sounding, introduced in Bluetooth 6.0, adds a complementary primitive that binds communication to physical proximity through encrypted phase-based ranging, defeating the relay-and-distance-fraud class of attacks that bypass higher-layer cryptography by forwarding packets between a legitimate device and a protected endpoint.\cite{skallak2023ble,khan2022capitalone} Composed with post-quantum session protocols at the application layer, these transport-level safeguards ensure that neural data, which implicitly functions as a biometric identifier (\cref{neural_identity}), is not only cryptographically valid but also physically local. With post-quantum secure session protocols at the application layer, these transport-level safeguards ensure that neural data is not only cryptographically valid but also physically local, embedding spatial integrity into the security properties of the BCI cycle.

\subsubsection{Acquisition-Layer Hardening Against Signal Injection}

Defenses against radio-frequency injection at the analog acquisition stage divide into shielding (twisted-pair cabling, ferrite suppression, and Faraday enclosures providing graduated attenuation), bandpass filtering ahead of the ADC to reject out-of-band carriers before in-amplifier rectification, and active runtime detection of the statistical signatures EMI injection leaves at the analog front-end.\cite{zhang2020detection} Filtering weakens as adversaries select carriers whose demodulation products fall within the EEG band, and the BCI form factor is incompatible with the heavy shielding that would provide robust protection.\cite{giechaskiel2020} Filtering and runtime detection are therefore the more practical near-term countermeasures, though the latter has not yet been adapted to BCI hardware specifically.

\subsection{BCI-Specific Defense Architectures}

Defense frameworks specific to BCI threat models exist as research prototypes rather than deployed standards, with work spanning signal-layer command filtering,\cite{bonaci2014, bonaci2015} system-level information flow control, \cite{tarkhani2022} and EEG-specific adversarial training and defense surveys.\cite{chen2024abat, wu2023advsurvey} A ratified, comprehensive defense stack analogous to TLS for BCI data streams has not yet emerged.

\section{Conclusion}

It is widely considered impossible to guarantee that any modern digital system is completely secure, simply because increasing the functional capabilities of a system--and especially increasing their breadth--tends to bring about an exponential increase in the set of potential vulnerabilities.\cite{Ferguson_2015_cryptography_engineering_textbook} If for no other reason than the combinatorics that each new interaction or function introduces potential new attack vectors for each other element within a system to which it is linked, this exponential growth is unavoidable. Consequently, as the capabilities of BCI systems develop and expand one must be prepared for this same inevitability. 

However, that is not an excuse for the kind of laxity seen with the Internet-of-Things, transformer-based AI systems (particularly agentic systems, which are famously rife with security issues at time of writing), and the early internet. Instead, there is a legal and ethical necessity to hold neurosecurity to far higher standards than that of digital hardware and software at large. While such issues are not the focus of this paper, it cannot be overstated that neurosecurity standards must be exceptionally high, and every effort must be taken to provide absolute guarantees whenever possible. Additionally, security is an ever-evolving challenge. As neurotechnology becomes more widespread, the dynamic of white hat (beneficent) and black hat (maleficent) actors will extend to BCIs. Malicious actors will constantly innovate new attacks and identify new vulnerabilities, and it is thus the responsibility of developers of neurotechnology to be constantly both vigilent and innovative themselves in order for defenses to remain effective against attackers. 

An analogy given by Ferguson and Schneier in their book \textit{Cryptography Engineering} is that many security systems are built in a manner akin to a tent with a bank vault door affixed to the front, with developers arguing over what features to add to the door while ignoring the frailty of the tent walls.\cite{Ferguson_2015_cryptography_engineering_textbook} Our review of attack surfaces in this paper is inevitably not perfectly exhaustive, though we have endeavored to provide as thorough as possible a cataloging of the \textit{types} of threats which creators of neurotechnologies face. There has been a historical trend where security is treated as an afterthought in the setting of emerging technologies, only taken seriously after significant harms have occurred and slow-moving government regulations or consumer awareness force change. Users of BCIs and other neurotechnologies are entrusting the devices' creators with the safety, privacy, and integrity of their very brains and minds, and this trust must be taken extremely seriously by researchers and technology developers alike. There is a moral imperative that the field refuse to repeat the mistakes of past revolutions in computing technologies. 

Every capability has a corresponding vulnerability--and more often there will be many, many vulnerabilities for each capability of a device. It is the responsibility of everyone working on neurotechnologies to remain conscientious of security threats, and vigilant and rigorous in defending against them.

\printbibliography

\end{document}